\documentclass[aps,prd,preprint,superscriptaddress,showpacs,nofootinbib]{revtex4-2}
\usepackage{epsf,epsfig,graphics,graphicx}
\usepackage{verbatim,color,ulem}
\newcommand{\be}{\begin{equation}}
\newcommand{\ee}{\end{equation}}
\newcommand{\bea}{\begin{eqnarray}}
\newcommand{\eea}{\end{eqnarray}}
\newcommand{\ba}{\begin{array}}
\newcommand{\ea}{\end{array}}
\usepackage{amssymb,amsmath,amsthm,graphicx}
\usepackage[mathscr]{eucal}
\usepackage{enumerate,color,verbatim,multirow,comment}
\usepackage[caption=false]{subfig}

\begin{document}

\title{Decoherence by black holes via holography}
\author{Shoichi Kawamoto}
\email{kawamoto\_s@obirin.ac.jp}
\affiliation{College of Arts and Sciences,
J.~F.~Oberlin University, Tokyo, Japan}
\author{Da-Shin Lee}
\email{dslee@gms.ndhu.edu.tw} \affiliation{Department of Physics,
National Dong Hwa University, Hualien, Taiwan, R.O.C.}
\author{Chen-Pin Yeh}
\email{chenpinyeh@gms.ndhu.edu.tw} \affiliation{Department of
Physics, National Dong Hwa University, Hualien, Taiwan, R.O.C.}

\begin{abstract}
In this note, we reexamine decoherence effects in quantum field theories with gravity duals. The thought experiment proposed in \cite{DSW_22, DSW_23}, which reveals novel decoherence patterns associated with black holes, also manifests itself from the perspective of the boundary theory. In particular, we consider a moving mirror coupled to quantum critical theories characterized by a dynamical exponent $z$ that are dual to asymptotically Lifshitz geometries.
The interference experiment occurs on the boundary, where a superposition of two spatially separated quantum states of a mirror is maintained for a finite time $\tau_0$ before recombination. We find that the interaction with a quantum field at finite temperature, arising from the presence of a Lifshitz black hole, leads to a constant decoherence rate. In contrast, for the zero-temperature case corresponding to pure Lifshitz spacetime, the decoherence rate vanishes in the large-time limit $\tau_0 \to \infty$. Remarkably, in the zero-temperature regime, the decoherence exhibits a power-law decay at large $\tau_0$ as $z \rightarrow \infty$, a behavior reminiscent of the decoherence patterns seen in extremal black hole geometries.
In addition, we investigate the decoherence of one particle in an EPR pair constructed holographically. Our results indicate that causality plays a crucial role in determining whether the entanglement leads to the suppression of decoherence in the other particle.
\end{abstract}

\maketitle

\section{introduction}
There has been a longstanding debate over whether black hole evaporation violates the principles of quantum mechanics. A central challenge lies in the lack of consensus, even among those who accept the validity of quantum mechanics, on how information is recovered in Hawking radiation. The potential loss of quantum information is often associated with decoherence. In this context, the black hole information problem can be viewed as a form of persistent decoherence induced by the black hole on nearby quantum systems, thereby violating the unitarity of quantum evolution.
Quantum decoherence typically arises from interactions with an environment. Recently, Danielson, Satishchandran, and Wald (DSW) proposed a thought experiment demonstrating that quantum superpositions maintained outside a Killing horizon will decohere at a constant rate \cite{DSW_22,DSW_23}. Their argument primarily relies on causality. They showed that when long-range fields, such as electromagnetic or gravitational fields, are entangled with quantum superposition states, soft-mode radiation can carry information across the horizon into the black hole, leading to a steady decoherence rate.
This decoherence effect, like Hawking radiation, appears to be related to the global structure of spacetime horizons. It can be analyzed using quantum field theory in curved spacetime, where the past state (in-vacuum) and the future state (out-vacuum) may differ \cite{Gralla_24}. More recently, it has been shown that the DSW effect can also be understood locally as quantum decoherence in a thermal environment \cite{Biggs_24,Chen_24,DSW_24}. These findings suggest that black holes can be modeled semiclassically as quantum systems in thermal equilibrium.
Nevertheless, it remains an intriguing question how the DSW effect manifests in a fully quantum-mechanical description of black holes and how decoherence can be minimized \cite{DSW_25}. Extending beyond the semiclassical regime is challenging due to the limited tools available. In this short paper, we aim to investigate the DSW effect using the local method in a holographic framework, which is believed to offer a fully quantum description of black holes.

In parallel with the study of the DSW effect, we consider an interference experiment involving a particle (or a mirror) in a coherent superposition of two localized states. These states depart from the same initial point and later recombine after following two distinct, non-overlapping paths, $C_1$ and $C_2$, as illustrated in Fig.~\ref{path}. Similar interference experiments have been studied in \cite{Ford_93,Ford_97}, where the influence of vacuum electromagnetic-field fluctuations on electron coherence was analyzed. These studies showed that both the phase shift and decoherence effects can be expressed through double worldline integrals involving the commutator and anticommutator of gauge potentials in the vacuum state, which are related to the retarded and Hadamard Green's functions, respectively.
In the framework of open quantum systems \cite{Breuer}, environment-induced effects on the system can be systematically described using the influence functional formalism developed by Feynman and Vernon \cite{Feynman}, constructed from Green's functions defined in the in-in formalism \cite{SC}. One notable application of this approach is the analysis of quantum Brownian motion in a thermal bath \cite{CAD}. The influence functional generally admits a decomposition into real and imaginary components: the real part, involving the Hadamard function, encodes decoherence, while the imaginary part, associated with the retarded Green's function, corresponds to phase shifts \cite{Hsiang_06}. These two effects are connected via the fluctuation-dissipation theorem.

In this note, we reexamine the decoherence of a moving mirror interacting with strongly coupled quantum critical fields with dynamical critical exponent $z$, using the influence functional derived from holographic duality \cite{Yeh_15,mirror}. In the holographic setup, a particle or extended object (referred to as a mirror) is modeled by a string or brane propagating in an asymptotically Lifshitz geometry. The influence functional for the interference experiment, conducted at the boundary, is obtained from the on-shell action of the string or brane in the bulk. The environmental fields at zero and finite temperatures are dual to pure Lifshitz spacetimes and Lifshitz black holes, respectively. In the finite temperature case, we interpret the environmental decoherence as arising from the black hole in the gravitational dual. We find that the resulting decoherence pattern agrees with that predicted by the DSW effect, where the particle in a superposition state loses coherence at a constant rate, and the decay rate scales with the cube of the black hole temperature. At zero temperature, we also find the recovery of coherence in the adiabatic late-time limit, which is consistent with the flat-space results. Furthermore, we observe that in the zero temperature case, as we increase the dynamical critical exponent $z$, the coherence recovers more slowly. Moreover, the recovery of coherence becomes polynomial in time as $z$ approaches infinity, which mimics the behavior of an extremal black hole. Thus, Lifshitz holography provides a continuous spectrum of decoherence behaviors that connects the case of flat space to that of black holes, and it shows that, in some sense, coherence recovers most slowly in extremal black holes.
Through the weak-strong coupling duality in the holographic dictionary, these results thus generalize those in \cite{Biggs_24,Chen_24,DSW_24} to strongly coupled field theories. The method also paves the way for studying the decoherence of a fully quantum black hole by considering the finite $N$ correction in the holographic dictionary. However, this may require a concrete model of the dual quantum field theory (one candidate is described in \cite{Chu_24}). Instead, we illustrate a quantum correction from a finite 't Hooft coupling in the zero-temperature case.
On the gravity side, this involves expanding the on-shell string action to subleading order in the string tension.
Additionally, we investigate decoherence arising from Unruh effects in the holographic setting by considering accelerating string endpoints \cite{Xiao_08,Yeh_22_1}. Here we find a constant decoherence rate, analogous to that observed in finite-temperature environments. By comparing this to the case of a static string with two endpoints, we can investigate the influence of one particle on the decoherence of the other particle in the entangled pair.   

In the next section, we outline the setup of the interference experiment and identify the decoherence function and phase shift as derived from the influence functional.  In Section \ref{sec:decoh-from-lifshs}, we present the calculation of the holographic influence functional and use it to analyze the resulting decoherence effect.  Section \ref{sec:decoh-due-unruh} explores a holographic description of the Unruh effect and its implications for decoherence. 
Finally, we summarize our findings and conclusions in Section \ref{sec:concluding-remarks}.

\section{review of influence functional }\label{decoherence}

In this section, we briefly review the influence functional formalism, describing how environmental degrees of freedom affect a quantum particle. We also introduce phase shifts and decoherence in an interference experiment in which the particle is in a superposition of two spatially separated trajectories \cite{Hsiang_06}. This formalism can be naturally extended to holographic settings, as demonstrated in \cite{Son_09,Yeh_15}.
 
The environment-induced effects on the system (a particle with position variable $q$) can be captured by a linear coupling to an environmental field $F$, with the full Lagrangian given by
  \be
   L(q,F)=L_q[q]+L_F[F]+qF\, .
  \ee
The evolution of the density matrix of the system-environment is obtained using the time-evolution operator  $ U(t_f,t_i) $ as
\begin{equation}
\rho (t_f) = U(t_f, t_i) \, \rho(t_i) \, U^{-1} (t_f,t_i )\, .
\end{equation}
We consider the case in which the initial density matrix at time $t_i$ can be factorized as
\begin{equation}\label{initialcond}
    \rho(t_i)=\rho_{q}(t_i)\otimes\rho_{{F}}(t_i)\,,
\end{equation}
and the environmental field $F$ is in thermal equilibrium at temperature $T=1/\beta $, with the density matrix $\rho_{{F}}(t_i)$
\begin{equation}\label{initialcondphi}
    \rho_{F}(t_i)=e^{-\beta H_{F}}\,,
\end{equation}
where  $H_{F}$ is the Hamiltonian of the environment field. The zero-temperature limit corresponds to taking $T \rightarrow 0$.
The reduced density matrix for the system is obtained by tracing out the environmental degrees of freedom.
In the position basis, it can be expressed as
\begin{eqnarray}
   \langle q_f|\rho_r|\tilde{q}_f\rangle\equiv \rho_r({q}_f,\tilde{{q}}_f,t_f)&=&\int\!d{q}_1\,d{q}_2\left[\int^{{q}_f}_{{q}_1}\!\!\mathcal{D}{q}^+\!\!\int^{\tilde{{q}}_f}_{{q}_2}\!\!\mathcal{D}{q}^-\;\exp\biggl\{i\int_{t_i}^{t_f}dt\left(L_{q}[{q}^+]-L_{q}[{q}^-]\right)\biggr\}\right.\nonumber\\
            &&\qquad\qquad\left.\,\times\, \mathcal{F}[\,q^+,q^-]\right]\rho_{q}({q}_1,{q}_2,t_i)\,, \label{evolve}
\end{eqnarray}
using the so-called influence functional
\begin{equation}
    \mathcal{F}[\,q^+,q^-]=\exp\Big\{\mathcal{W}[\,q^+,q^-]+i\,\Phi[\,q^+,q^-]\Bigr\}\,.
\end{equation}
Here, the real functionals $\mathcal{W}$ and $\Phi$ are given by
\begin{align}
    \Phi[\,q^+,q^-]&=\frac{1}{2}\int\!dt \!\!\int\!dt'\Bigl[\,q^+(t)-q^-(t')\Bigr]G_{R}(t-t')\Bigl[\,q^+(t) + q^-(t')\Bigr]\,,\label{phase}\\
    \mathcal{W}[\,q^+,q^-]&=-\frac{1}{2}\int\!dt \!\!\int\!dt'\Bigl[\,q^+(t)-q^-(t')\Bigr]G_{H}(t-t')\Bigl[\,q^+(t) - q^-(t')\Bigr]\,,\label{decoherence}
\end{align}
where the retarded Green's function $G_{R} (t-t')$ and the Hadamard function $G_{H} (t-t')$ of the environmental field are defined, respectively, as
\begin{eqnarray}
    G_{R} (t-t')&=&-i\,\theta(t-t')\,\bigl<\left[{{F}} (t),{{F}} (t')\right]\bigr>\,,\label{commutator}\\
    G_{H} (t-t')&=& \frac{1}{2}\,\bigl<\left\{{{F}} (t),{{F}} (t')\right\}\big>\,.
    \label{anticommutator}
\end{eqnarray}
They obey the fluctuation-dissipation relation
  \be
  G_H(\omega)=- \coth \bigg[\frac{\omega}{ 2T}\bigg] \operatorname{Im}G_R(\omega)
  \, \label{FDR_T}
  \ee
in frequency space. At zero temperature $T\to 0$, the relation reduces to
\be \label{FDR_zeroT}
  G_H(\omega)=  -\operatorname{sgn}(\omega) \operatorname{Im}G_R(\omega) \, .
  \ee

\begin{figure}
\centering
    \scalebox{0.5}{\includegraphics{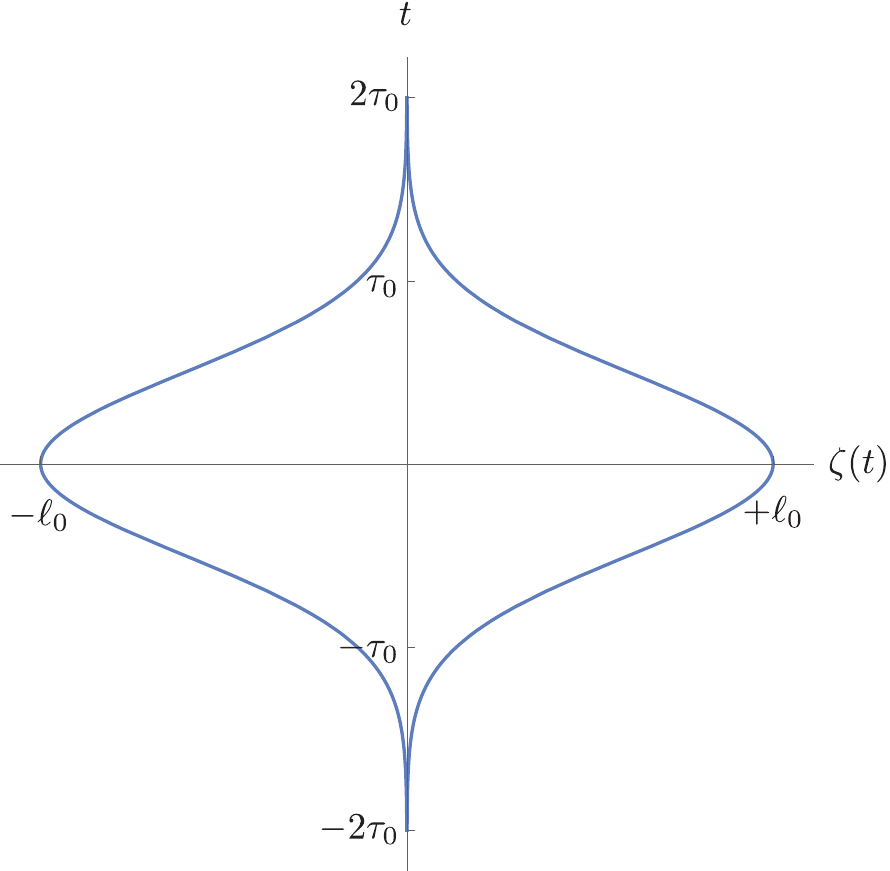}}
\caption{The spacetime paths of the wavepackets for an
interference experiment.}\label{path}
\end{figure}

We now apply the influence functional formalism outlined above to the interference experiment described below. Suppose that the initial state of the particle is a coherent superposition of two localized states, with the initial density matrix given by
\begin{eqnarray}
    \rho_q(t_i)=( |\psi_1(t_i)\rangle+|\psi_2(t_i)\rangle) (\langle\psi_1(t_i)|+\langle\psi_2(t_i)|)
\end{eqnarray}
where $|\psi_1(t)\rangle$ and $|\psi_2(t)\rangle$ represent wave packets propagating along the two non-overlapping paths ${C}_1$ and ${C}_2$ separately, as shown in Fig.~\ref{path}.  We assume that the wave packet is sharply peaked in both position and momentum, and that wave-function spreading is negligible. Under this approximation, the leading contribution to decoherence can be obtained by evaluating the influence functional along prescribed classical trajectories. The diagonal components of the reduced density matrix then become
\begin{equation}
\label{interf}
    \rho_r({q}_f,{q}_f,t_f)=\bigl|\psi_1({q}_f,t_f)\bigr|^2+\bigl|\psi_2({q}_f,t_f)\bigr|^2+2\,e^{\mathcal{W}[\,\bar{q}^+,\bar{q}^-]}\;\operatorname{Re}\left\{e^{i\,\Phi[\,\bar{q}^+,\bar{q}^-]}\psi_1^{\vphantom{*}}({q}_f,t_f)\,\psi_{2}^{*}({q}_f,t_f)\right\}\,,
\end{equation}
where the $\mathcal{W}$ and $\Phi$ functionals are evaluated along the classical trajectories ${C}_1=\bar{q}^{+}$ and ${C}_2=\bar{q}^{-}$. The last term in Eq.~\eqref{interf} gives the interference effect; $\Phi$ gives the phase shift in the interference pattern, while $\mathcal{W}$ determines the strength of the interference. When $\mathcal{W}$ is large and negative, the interference pattern disappears. Thus, we refer to $\mathcal{W}$ as the decoherence functional, which characterizes the decoherence effect.

Here we consider the prescribed trajectories $\bar{q}^{\pm}= \pm \zeta (t) $ for the localized states. The path function $\zeta(t)$ is chosen to take the form
\begin{equation}\label{paths}
    \zeta(t)={ \ell}_0\,e^{-\frac{t^2}{\tau_0^2}}\,,
\end{equation}
where $2\ell_0$ is the effective path separation and $2\tau_0$ is the effective flight time. The decoherence functional from \eqref{decoherence} is obtained as
   \be \label{W}
   \mathcal{W}=-  \int\frac{d\omega}{2\pi}\; G_H(\omega) \bigg\vert \int
   dt \, \zeta(t) \, e^{-i\omega t} \bigg\vert^2 <0\, .
   \ee
Since these two prescribed classical trajectories are symmetric with respect to the initial position, we have $q^{+}(t)+q^{-}(t)=0$ and therefore no phase shift appears; i.e., $\Phi =0$ from \eqref{phase}. In the following sections, we will use the holographic influence functional to evaluate $\mathcal{W}$ and interpret the decoherence effect as a result of the gravity dual theory.   

\section{Decoherence from Lifshitz spacetimes}\label{sec:decoh-from-lifshs}
In this section, we consider an $n$-dimensional moving mirror coupled to an environmental field in a theory of $(d+1)$-dimensional quantum critical points \cite{mirror,Yeh_18_1}, which exhibits the following scaling symmetry:
  \be\label{E:ernfd}
  t\rightarrow\mu^zt \, ,\qquad\qquad x\rightarrow\mu x \,,
  \ee
where $z$ is called the dynamical critical exponent. The gravitational dual of this quantum critical theory is described by the Lifshitz geometry with the following asymptotic metric\footnote{To obtain this metric, we need to couple gravity to background form fields (see \cite{squeezed}, for example). This then raises the question of whether the decoherence effect considered below may have an extra contribution from the interaction of the string (or brane) with the form fields. In the model we consider, the string (or brane) couples to the metric only, and in the probe limit, there is no coupling of the fluctuations to the form fields. We leave it for a future project to see whether different models produce different decoherence patterns.} near the boundary ($r\to r_b\gg 1$),
   \be
   \label{bmetric}
   ds^2=-\frac{r^{2z}}{L^{2z}}dt^2+\frac{1}{r^2}dr^2+\sum_{i=1}^{d-1}\frac{r^2}{L^2}dx_idx_{i}
   \, ,
   \ee
where the scaling symmetry \eqref{E:ernfd} is realized as an isometry of this metric \cite{Tong_12,Taylor_08}. In the following, the curvature radius $L$ is set to one (we will restore it in the formulas for stringy corrections).  In the gravity picture, the mirror (or the particle in the case $n=0$) is described by an $(n+1)$-dimensional probe brane (or a probe string when $n=0$) in the Lifshitz geometry. We consider the decoherence effects induced by zero-temperature and finite-temperature environments, which are dual to the pure Lifshitz geometry with the metric \eqref{bmetric} and the Lifshitz black hole geometry, respectively. In both cases, the dynamics of probe branes (or strings) is described by world-volume theories, with action $S=-T_{n+1}\int d^{n+2}\vec{\sigma}\sqrt{-\det h_{ab}}$, where $T_{n+1}$ is the brane tension and $h_{ab}$ is the induced metric on the brane (or string, if $n=0$) with the world-volume coordinates $\vec{\sigma}$. According to the holographic prescription in the so-called large-$N$ approximation, the influence functional is identified with the classical on-shell world-volume action for small fluctuations $X$ of the probe brane (or particle) in the maximally extended Lifshitz black hole background
  \be
  \label{holoF}
  \mathcal{F}[q^+,q^-]=\exp (S_\text{on-shell}(X^+,X^-)) \,,
  \ee 
 where $X^+$ and $X^-$ are two branches of the classical solution in the two-sided Lifshitz black hole with the following boundary conditions: (1) at the boundary $r=r_b$, they match the prescribed values $q^+$ and $q^-$ separately; and (2) near the horizon $r=r_h$, they satisfy the infalling boundary conditions. The zero-temperature holographic influence functional is obtained by taking the $T \to 0$ limit of the finite-temperature result (see \cite{Yeh_15} or \cite{Son_09} for details).    

\subsection{Zero Temperature}

In the zero-temperature case, the probe brane (or string) has static solutions with one endpoint fixed at the boundary. The quantum fluctuation of the string endpoint (which is interpreted as the motion of a Brownian particle from the boundary point of view) is holographically described by small perturbations around the static configuration. The action of these perturbations up to quadratic order is given by \cite{Yeh_15}
  \bea
  S\simeq - \frac{T_{n+1} S_n}{2}\int dr \, dt \,
\bigg( r^{z+n+3}  \partial_{r} X\partial_{r} X-
\frac{\partial_{t}X\partial_{t}X}{ {  r^{z-n-1}}}\bigg) \, ,
  \label{s_zero}
  \eea
where $S_n$ is the volume of the brane, and $X(t,r)$ represents one of the transverse displacement modes of the brane around its static configuration, parameterized by the metric time and radius in \eqref{bmetric} (in the case $n\geq 1$, we assume the mirror is rigid, so the perturbations are independent of the coordinates parallel to the direction of the brane). Since the perturbations in different directions decouple in the quadratic action, we consider only one of them.  Following the prescription developed in \cite{Yeh_15} or \cite{Son_09}, the related Green's functions of the boundary fields that couple to the mirror can be constructed holographically by evaluating the on-shell action in \eqref{s_dbi} using the solutions of $X(t,r)$ satisfying the appropriate boundary conditions, as described in Eq.~\eqref{holoF}. Using this approach, the zero-temperature retarded Green's function for $\omega>0$  was found to be \cite{Tong_12,mirror, Lee:2019adw},
  \be
\label{G_R} G_R(\omega)=- T_{n+1} S_n \,  {\omega \,
r_b^{n+2}}\frac{H^{(1)}_{\frac{\alpha}{2}-1}(\frac{\omega}{zr_b^z})}{H^{(1)}_{\frac{\alpha}{2}}(\frac{\omega}{zr_b^z})}
\,,
  \ee
where $H^{(1)}_{\nu}(x)$ is the Hankel function of the first kind, $r=r_b$ is the UV cutoff for the location of the brane, and $\alpha\equiv1+(n+2)/z$. To examine the late-time behavior of the fluctuations, we consider the small $\omega$ expansion of $G_R(\omega)$,
 \be \label{G_R_approx}
  G_R (\omega)={ m (i\omega)^2+\gamma (-i\omega)^{\alpha}}+ {\cal{O}}( \omega^2/r_b^{2z}) \, ,
  \ee
 where
  \be \label{m_mu_n}
   m =\frac{T_{n+1} S_n}{z(\alpha-2)r_b^{z(2-\alpha)}},
  \ee
and
 \be \label{gamman} \gamma =\frac{T_{n+1}
S_n}{(2z)^{\alpha-1}}\frac{\Gamma(1-\frac{\alpha}{2})}{\Gamma(\frac{\alpha}{2})}
\, .
 \ee
Thus, the interaction with the environmental field not only generates an effective mass $m$ for the mirror that depends on the cutoff $r_b$, but also produces a damping effect with the coefficient $\gamma$. The friction coefficient $\gamma$ is independent of $r_b$, which is important for stabilizing the dynamics of the mirror in a fluctuating environment. Since $\alpha >1$, this corresponds to the so-called supraohmic dissipation. For example, in the case of a relativistic two-dimensional mirror, one finds $\alpha=5$. The zero-temperature Hadamard function for $\omega>0$ was found to be
 \be \label{G_0_H}
 G_H(\omega)=\frac{2 z}{\pi} r_b^{n+2+z} \frac{T_{n+1} S_n} {J^2_{\frac{\alpha}{2}}(\frac{\omega}{zr_b^z})+Y^2_{\frac{\alpha}{2}}(\frac{\omega}{zr_b^z})}
  \, , \ee
where $J_{\nu}(x)$ and $Y_{\nu}(x)$ are Bessel functions of the first and second kind, respectively.
In the small $\omega$ limit, the Hadamard function can be approximated as
\be \label{G_0_H_approx}
 G_H(\omega)=\frac{\pi \, T_{n+1} S_n }{ (2 z)^{\alpha-1} \, \Gamma^2 (\frac{\alpha}{2})} \, \omega^{\alpha} + {\cal{O}}( \omega^{\alpha+2}/r_b^{z(\alpha+2)}) \, .
\ee 
Notice that the obtained retarded Green's function and the Hadamard function obey the zero-temperature fluctuation-dissipation theorem \eqref{FDR_zeroT} \cite{Son_09,Boer_14}. 
Thus, in the small-$\omega$ limit, the leading effect on the decoherence functional in \eqref{W}
is
    \be \label{W_zero_T}
\mathcal{W}\simeq-\frac{\pi \, T_{n+1} S_n \, \Gamma (\frac{1+\alpha}{2})}{ ( 2)^{(1+\alpha)/2} \, \Gamma^2 (\frac{\alpha}{2})}\,
\frac{\ell_0^2}{(\tau_0 z)^{\alpha-1}}+ {\mathcal O}(1/r_b^{2z(\alpha+2)}\tau_0^{2(\alpha+2)}) \,.
    \ee
Since $\alpha >1$, in the long-time limit $\tau_0 \rightarrow \infty$, we have $\mathcal{W}\rightarrow 0$. This implies that the decoherence of the system with supraohmic dissipative dynamics can be vanishingly small for adiabatic motion of the quantum states.  This is consistent with the findings in \cite{Chen_24}, where the decoherence effect on the interference experiment vanishes in the adiabatic limit when the environment exhibits supraohmic friction. 
While in their case the environment consists of the electromagnetic fluctuations (and gives $\alpha=3$), in our case, the environment is described by the gravity dual of pure Lifshitz geometry, allowing any value of $\alpha>0$. As we will see in the next section, when the environment is dual to the Lifshitz black holes, the system exhibits ohmic friction, and this leads to a constant decoherence rate in the large $\tau_0$ limit. This is consistent with the local interpretation of the DSW effect in \cite{Chen_24,Biggs_24}. It is also interesting to note that, since we have $\alpha=1+\frac{n+2}{z}$, the decoherence effect vanishes more slowly for larger values of $z$. The friction term in Eq.~\eqref{G_R_approx} actually approaches an ohmic one as $z\to\infty$. However, Eq.~\ref{W_zero_T} does not yield a constant decoherence rate in this limit. Instead, in the large-$\tau_0$ and $z\to\infty$ limit, we find
  \be
  \mathcal{W}\propto 1-\frac{(n+2)}{z}\ln(\tau_0)+... \,.
  \ee
We thus observe a logarithmic decay of the decoherence rate, and this can be compared with the decoherence effect caused by an extremal black hole at zero temperature, as found in \cite{Gralla_24}. It is also noted in \cite{Chen_24} that ohmic friction is not sufficient to produce the DSW effect, which gives a constant decoherence rate; a thermal environment is also required. Before we investigate the holographic counterpart of this finite temperature result, we discuss the stringy correction to the zero-temperature decoherence behavior.

\subsection{Stringy Correction}

In this section, we consider the case of the string ($n=0$) with the string tension $T_1\equiv \frac1{2\pi\alpha'}$. To estimate the $\alpha'$ correction to the decoherence rate, we consider the quartic correction terms to the world-volume action in (\ref{s_zero}), which take the form\footnote{At quartic order, there are also terms that mix the fluctuations in different directions. For simplicity, we only consider the fluctuation in one direction, since the purpose of this section is not to completely calculate the $\alpha'$-corrections, but just to demonstrate the possibility of the correction and the way it appears in the holographic method.} (see \cite{mirror} for example). We now define $\partial_r\equiv ~'$ and $\partial_t\equiv ~\dot{}$, and restore the curvature radius $L$.
  \be
  S_{quartic}=\frac{L^2}{16\pi\alpha'}\int drdt \left( r^{z+7}X'^4+2r^{-z+5}X'^2\dot{X}^2+r^{-3z+3}\dot{X}^4 \right)\, . 
  \ee
Then, including the quadratic terms in (\ref{s_zero}), we have the following equation of motion for the string fluctuation,
 \be
 \label{neom}
 \partial_r(r^{z+3}X')-r^{-z+1}\partial_t^2X-\frac12\partial_r(r^{z+7}X'^3)+\frac12\partial_r(r^{-z+5}X'\dot{X}^2)+\frac12\partial_r(r^{-z+5}\dot{X}X'^2)-\frac12r^{-3z+3}\partial_t\dot{X}^3=0 \, .
  \ee
Without the non-linear terms in (\ref{neom}), we have the following mode expansion that is suitable for the Hadamard function (see \cite{mirror} for details)
 \be
 X(t,r)=\int d\omega U_{\omega}(r)(a_{\omega}e^{-i\omega t}+a_{\omega}^{\dagger}e^{i\omega t})
 \ee 
 In this case, the Hadamard function in Fourier space is given by
  \be
  \label{fdr}
  G_H(\omega)=2\pi U^2_{\omega}(r_b)G_R(\omega)G_R(-\omega)
  \ee
 Now we look for a correction to the string fluctuation $X(t,r)$ coming from the nonlinear terms in (\ref{neom}). In \cite{mirror} we have the exact solution for the function $U_{\omega}(r)$, which gives the Hadamard function related to the retarded Green function, $G_R(\omega)$ in (\ref{G_R}) by the fluctuation-dissipation relation, $G_H(\omega)=\mbox{sign}\, \omega \; \mbox{Im}\, G_R(\omega)$. Here, we are only interested in the late-time and large-$r$ behavior of $X(t,r)$. So we consider $U_{\omega}(r)$ in the limit $\omega\ll r$, and it gives (we also let $a_{\omega}\rightarrow \omega^{-1/2}$ by dimensional analysis)
  \be
  X(t,r)\simeq \frac{\sqrt{\alpha'}}{L}r^{z-2}t^{1-\frac1z}
  \ee 
Taking this as a zeroth-order solution to the equation (\ref{neom}), we find that, to order $\frac{\alpha'}{L^2}$, a perturbative solution (valid for small $\omega$ and large $r$) reads
  \be
  U_{\omega}(r)\propto r^{z-2} \omega^{\frac1z-\frac32}+C(z)\frac{\alpha'}{L^2}r^{3z-4}\omega^{-\frac72+\frac3z} \, ,
  \ee
  where $C(z)$ is some constant depending on $z$. From (\ref{fdr}), this gives the Hadamard function 
   \be
   G_H(\omega)\propto \omega^{1+\frac2z}+2C(z)\frac{\alpha'}{L^2} r_b^{2z-2} \omega^{-1+\frac4z}
   \ee
Furthermore, from (\ref{W}), we obtain the stringy-corrected decoherence functional
 \be
 \label{stringy}
  \mathcal{W}\propto \tau_0^{-\frac2z}+2C(z)\frac{\alpha'}{L^2}r_b^{2z-2}\tau_0^{2-\frac4z}
  \ee
Thus, for $z<2$, the correction term is not important at late times. When $z>2$, the correction can be important\footnote{This qualitative change of dissipation behavior across $z=2$ can also be seen in the study of entanglement entropy \cite{Lee:2019adw}. Notice that the expression for the Green's functions is not valid for $z=2$. For the discussion of the valid range for $z$, see also \cite{Lee:2019adw}. }, and it may help or suppress the recovery of coherence depending on the sign of $C(z)$ (However the perturbation theory breaks down as the growing correction term is of the same order as the leading term. In equation (\ref{stringy}), this happens at the time scale $\simeq (\frac{\alpha'}{L^2})^{-\frac{z}{2z-2}}r_b^{-z}$ ). Since the analysis in this section is meant to be an order-of-magnitude estimate, we do not have information about the sign of $C(z)$. Note also that this is not the only term of order $\frac{\alpha'}{L^2}$. There may also be other contributions to the decoherence of this order that come from the corrections to $G_R(\omega)$ and the normalization of $U_{\omega}(r)$, which are not included in the current work.

\subsection{Finite Temperature}

In this section, we consider the decoherence effect arising from a finite-temperature environment that is dual to a Lifshitz black hole.
The background metric for a Lifshitz black hole in $(d+1)$ dimensions is
  \be
  \label{lifshitz bh}
  ds^2=-r^{2z}f(r)dt^2+\frac{dr^2}{f(r)r^2}+r^2d\vec{x}^2 \, .
  \ee
The function $f(r)$ satisfies  $f(r)\rightarrow1$ as $r\rightarrow\infty$, and
$f(r)\simeq c(r-r_h)$ near the black hole horizon $r_h$, with $c=({d+z-1})/{r_h}$. 
The temperature of the black hole, which corresponds to the temperature of the boundary field theory, is 
 \be 
 \label{r_h}\frac1T=\frac{4\pi}{d+z-1}\frac1{r_h^z}\, . 
 \ee
The world-volume action for small fluctuations of the probe brane/string becomes
 \bea
  S^{(T)} \simeq - \frac{T_{n+1}S_n}{2}\int dr \, dt \,
\bigg( r^{z+n+3}  f(r) \partial_rX\partial_rX-
\frac{\partial_tX\partial_tX}{ {  r^{z-n-1} f(r)}}\bigg) \, ,
  \label{s_dbi}
  \eea
The relevant Green's functions of the environmental field, which couple to the mirror (or particle) in the interference experiment, can also be obtained using the holographic influence functional method \cite{mirror}. In the low-frequency and high-temperature approximation, the retarded Green's function has been found, in the regime $\frac{r_h}{r_b}\gg\frac{\omega}{r_b^z}$, to be
 \be
  G^{(T)}_{R}(\omega)=-\gamma_{T}(z)\, i\,\omega+m_{T}(z) (i\omega)^2+\mathcal{O}(\omega^3)\,,
  \ee
 where
 \be \label{m_gamma_T} m_{T}(z)=\frac{T_{n+1}S_n}{r_b^{z-n-2}}\biggl\{
\frac1{n+2-z}+\bigg(\frac{r_h}{r_b}\bigg)^{2n+4}\Bigl[(n+2+z)-\kappa
r_b^{z+n+2}\Bigr]\biggr\}\, , \, \gamma_{T}(z)=T_{n+1}S_n r_h^{n+2}
\,. \ee
Here $\kappa$ is some order-one integration constant that depends on the explicit form of $f(r)$. 
The mass $m_{ T}$ and the damping coefficient $\gamma_{T}$ depend on the temperature through the black hole temperature \eqref{r_h}. Since the damping term has linear dependence on $\omega$, the dissipation dynamics are ohmic. Through the fluctuation-dissipation relation \eqref{FDR_T}, the Hadamard function at high $T$ is obtained to leading order in the small-$\omega$ expansion as
    \be \label{G_H_T}
    G^{(T)}_H(\omega)\simeq
    2T\gamma_{T}(z)= {2T} \cdot
T_{n+1}S_n
\left(\frac{4\pi
    T}{d+z-1}\right)^{\frac{n+2}{z}} \, .
    \ee
Using Eq.~\eqref{W}, we obtain the finite-temperature decoherence function
    \be\label{W_T_z}
 \mathcal{W}_T\simeq - 
T_{n+1}S_n \ell_0^2 \tau_0 
\sqrt{\frac{\pi}{2}}T\left(\frac{4\pi T}{d+z-1}\right)^{\frac{n+2}{z}}
    \ee
 which yields a constant decoherence rate as expected for the DSW effect. Note also that in the case of a relativistic particle ($z=1$ and $n=0$), we obtain the $T^3$ scaling of the decoherence rate, in agreement with the result in \cite{DSW_22,DSW_23}.

\section{decoherence due to Unruh effect}\label{sec:decoh-due-unruh}

As demonstrated in \cite{Chen_24}, the DSW effect can also be understood through the decoherence of an accelerating charged particle arising from electromagnetic field fluctuations associated with the Unruh effect. In this section, we first review their argument and then illustrate a parallel analysis in the holographic setting.

\subsection{decoherence of an accelerating charged particle}

The dynamics of a charged particle with charge $Q$ and trajectory $x^\mu$ obeys the relativistic Abraham-Lorentz-Dirac (ALD) equation, where the radiation reaction force term can be written using the proper time $\tau$ as
\be
f^\mu=\frac{Q^2}{6 \pi} \left[\frac{d^3 x^\mu}{d\tau^3}-\frac{ d x^\mu}{d \tau}\left( \frac{d^2 x^\nu}{d\tau^2}
\frac{d^2 x_\nu}{d\tau^2} \right) \right]\, .
\ee
Consider a small perturbation $\epsilon^\mu(\tau)$ around the uniformly accelerating trajectory $x^\mu=z^\mu+\epsilon^\mu$, where the unperturbed trajectory satisfies $\ddot z^\mu=a^2 z^\mu$ and $\epsilon^\mu z_\mu =0$. The perturbed ALD equation becomes
\be
f^\mu=\frac{Q^2}{6\pi} \left[\frac{d^3 \epsilon^\mu}{d\tau^3}-a^2 \frac{d \epsilon^\mu}{ d \tau}\right] \,,
\ee
which leads to the damping coefficient in frequency space $\omega$ as
\be
\gamma=\frac{Q^2}{6\pi}( a^2 \omega+\omega^3)\, .
\ee
Thus, for a uniformly accelerating charge, the dissipative dynamics are ohmic and give rise to a constant decoherence rate, as discussed earlier. However, for a static charge in a zero-temperature environment, the damping effect becomes supraohmic, and from \eqref{W_zero_T} we see that the decoherence functional vanishes like $\tau_0^{-2}$ for adiabatic motion of the particle as $\tau_0\rightarrow \infty$. It is important to note that being in a non-inertial frame is crucial for the DSW effect. We can see this in the case of a static charge in a thermal environment. From the finite-temperature fluctuation-dissipation relation \eqref{FDR_T} and Eq.~\eqref{W}, we find that in this case the decoherence functional also vanishes in the adiabatic limit as $\tau_0^{-1}$.

\subsection{decoherence of holographic EPR pairs due to the Unruh effect}

One way to study Unruh effects in holography is by considering the exact string solution in AdS space with its two endpoints accelerating in opposite directions. This model was proposed in \cite{Xiao_08} to study an entangled heavy quark-antiquark pair (an EPR pair) in the boundary theory.  
Here, we explore the decoherence effect on one of the entangled particles in this setup.
The $AdS_{d+2}$ metric in the Poincar\'e coordinates is given by \eqref{bmetric} with $z=1$. Unlike in the previous sections, we consider only the world-volume theory for a string (namely, only $n=0$). Unlike the static string solution as considered in the previous section, where there is only one end located on the boundary, we consider an exact time-dependent string solution \cite{Xiao_08} 
  \be
  \label{ws}
  x_b(t,r)=\pm\sqrt{t^2+b^2-\frac{1}{r^2}} \,,
  \ee
where $b$ is a real constant and $x_b$ can be any direction chosen from the $x_i$ in \eqref{bmetric}, say $x_1$. In this solution, the trajectory of the endpoints at $r=r_b$ describes the motion of two particles along the $x$-direction with uniform deceleration $\frac{1}{b}$. They head toward each other from $x=\pm\infty$ when $t=-\infty$, subsequently stop at $x=\pm b$ when $t=0$, and then turn around in the opposite directions, moving away from each other. It is interesting to note that the induced metric on the worldsheet \eqref{ws} is nothing but a two-dimensional two-sided AdS black hole with Hawking temperature $T_b=\frac{1}{2\pi b}$, which is exactly the Unruh temperature for a detector of constant acceleration $\frac{1}{b}$ \cite{shock}.
  
To obtain the holographic influence functional, we consider small fluctuations around the string in a transverse direction, say $x_2$, with the embedding $X^{\mu}=(t,r ,x_b(t,r),q(t,r),0,\cdots, 0)$ in \eqref{bmetric}, where $q(t,r)$ is small compared to the background solution $x_b$.
The quadratic action for the fluctuation $q(t,r)$ is
  \be
  S_{qc}=-\frac{T_1}{2}\int dt dr \frac{1}{r^2}\left(\Big(1-\frac{r^2}{b^2}\Big)\partial_rq\partial_rq-\frac{\partial_tq\partial_tq}{1-\frac{r^2}{b^2}}\right)  \,
  \ee
where $T_1$ is the string tension. 
Using the holographic prescription described in the previous section, we obtain, in the small-frequency limit $\omega\ll \frac1b\ll r_b$, the retarded Green's function for the boundary field coupled to one of the entangled particles \cite{Yeh_22_1}
 \be
 G_R(\omega)
=-i\gamma_b\omega-M_b\omega^2+... \,,
 \ee
with $\gamma_b=\frac{T_1}{b^2}$ and $M_b={ T_1}{r_b}$. Using the fluctuation-dissipation relation \eqref{FDR_T} with the Unruh temperature $T_b=\frac1{2\pi b}$ and \eqref{W}, we obtain the decoherence functional
  \be
   \mathcal{W} =- \sqrt{\frac{\pi}{2}}\gamma_b \ell_0^2 \tau_0 T_b \propto -T_b^3\tau_0\, ,
   \ee
which also gives a constant decoherence rate with the right scaling in temperature, consistent with the DSW effect.

\subsection{decoherence of static EPR pairs}
For comparison with the decoherence of a static charge, we also consider a case where the two ends of the string are stationary with respect to each other.  The classical equation of motion for the static string with its two ends fixed at the boundary $r=r_b$ has a two-branch solution, joined at the turning point $r=r_0$\cite{Yeh_22_1}, 
   \be
    \label{zeroT}
    x_0(r)=\pm\int_{1/r_0}^{1/r}\sqrt{\frac{y^4}{\frac{1}{r_0^4}-y^4}}dy \, ,
    \ee  
where $x_0$ stands for one of $x_i$ directions in \eqref{bmetric}. In the limit $r_0 \ll\omega\ll r_b$, using the holographic influence functional method, we find the retarded Green's function \cite{Yeh_22_1}
  \be
  G_R(\omega)\simeq -M_0 \omega^2-i\gamma_0\omega^3+...
  \ee
  where $M_0=T_1 r_b$ and $\gamma_0= T_1$. Using the zero-temperature fluctuation-dissipation relation and Eq.~\eqref{W}, we obtain the decoherence functional, which decays as $\tau_0^{-2}$ in the large-$\tau_0$ limit. This can be compared with the case of a static charge in a zero-temperature environment. 
 
  \begin{figure}[h]
\subfloat[The static string worldsheet]{\label{fig1}\includegraphics[scale=0.42]{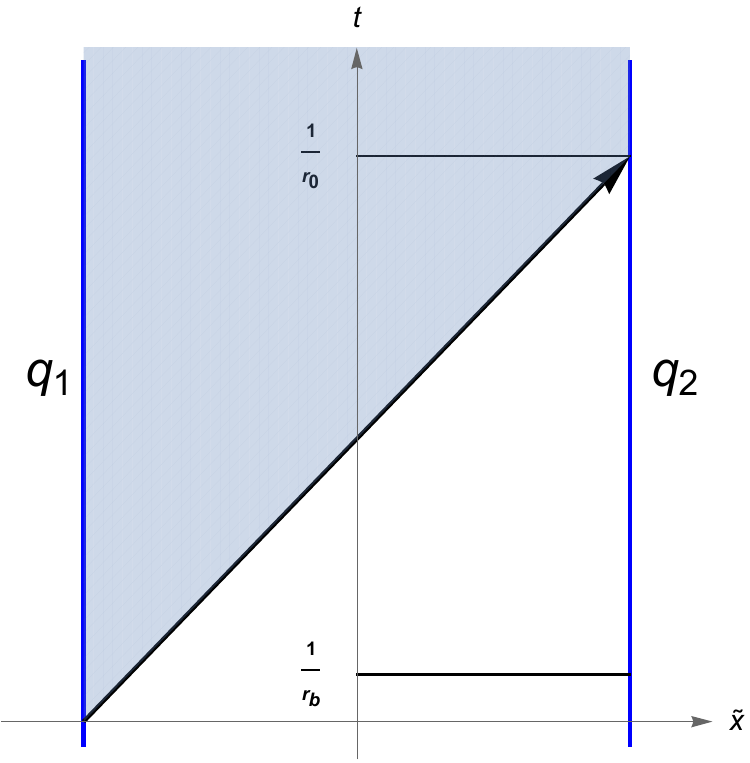}} \qquad \qquad \qquad \subfloat[Accelerating string worldsheet]{\label{fig2}\includegraphics[scale=0.42]{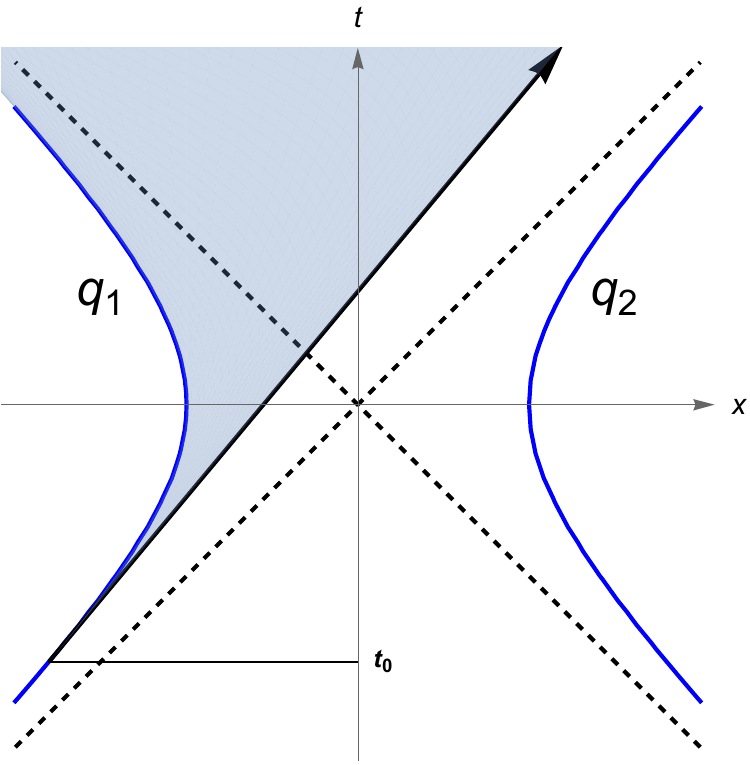}}
\caption{The causal diagrams for the induced metrics on the static string worldsheet in (\ref{zeroT}) (parameterized by $t$ and the conformal coordinate $\tilde{x}$) and the accelerating string worldsheet in (\ref{ws}) (parameterized by $t$ and $x$), respectively. In both diagrams, $q_1$ and $q_2$ are the trajectories of the string endpoints. Two large arrows are light rays emitted from $q_1$ at the times $t=0$ and $t=t_0$, respectively.  The shaded regions include those points in the future that are causally connected to the events when the light rays were emitted. Thus, in the static case, the interference experiment on $q_2$ can be affected by $q_1$ after $t\sim\frac1{r_0}$. In the accelerating case, the two endpoints are causally separated by the bifurcate horizons (the dashed lines), so there is no effect from $q_1$ on the interference experiment performed on $q_2$.}
\label{causality}
\end{figure} 
However, as pointed out in \cite{Yeh_22_1}, the retarded Green's function in this case becomes real in the limit $\omega\ll r_0\ll r_b$. The fluctuation-dissipation relation then implies that the Hadamard function vanishes in this limit. Hence, the decoherence effect in the interference experiment also vanishes. 
This seems inconsistent with the polynomial decay of the decoherence for a static charge in a zero-temperature environment. This discrepancy can be explained by noticing that here the particle in the interference experiment is entangled with the other particle. Since $\frac{1}{r_0}$ is the length scale for the separation of the EPR pair, in the first limit $r_0 \ll\omega\ll r_b$, the time scale is short compared to the separation, but large compared to the cutoff scale $\frac{1}{r_b}$ (that is, $\frac1{r_b}\ll t\ll \frac1{r_0}$). In this time region, the other particle remains causally disconnected from the one in the interference experiment, and the decoherence arises solely from the environment. In the second limit $\omega\ll r_0\ll r_b$ (which corresponds to the time region $t\gg\frac1{r_0}$), the time scale we consider allows the other entangled particle to affect the one in the interference experiment, and that turns out to suppress the decoherence at late times. 
Thus, we observe that causality plays an important role in the decoherence effect on one particle in the entangled pair. We illustrate this in Fig.~\ref{causality}. This also yields a phenomenon analogous to the electron decoherence from zero-temperature electromagnetic field fluctuations in the presence of a conducting plate. The effect of the conducting plate can be treated in terms of the image charges, as in \cite{Hsiang_06}.

\section{Concluding remarks} \label{sec:concluding-remarks}

In this note, we investigate the decoherence effects on a moving mirror (or particle) influenced by quantum critical theories with dynamical exponent $z$, using holography. The holographic dual is given by the Lifshitz geometry. We study the nonequilibrium motion of the mirror (or particle) using the holographic influence functional, whose real part involves the retarded Green's function $G_R$, responsible for the phase shifts, while its imaginary part involves the Hadamard function $G_H$ for decoherence. The Green's functions $G_R$ and the $G_H$ obtained this way also obey the fluctuation-dissipation theorem.
We study the decoherence effect in an interference experiment, where the superposition of two mirror (or particle) states is kept spatially separated for a time scale $\tau_0$ before being recombined. For pure Lifshitz spacetime, which is dual to the zero-temperature environment, the retarded Green's function obtained in this way gives supraohmic dissipation, and it induces a decoherence functional that vanishes in the large-$\tau_0$ limit for any finite $z$. However, as $z$ becomes larger, the decoherence vanishes more slowly. Moreover, as $z\rightarrow \infty$, both Green's functions  $G_R$ and $G_H$ have a linear dependence on $\omega$, and this results in a $\ln \tau_0$ dependence in the decoherence influence functional. This can be compared with the DSW effect in the case of extremal black holes. Thus Lifshitz holography provides a continuous spectrum of coherence-recovering rates that connects the flat-space case to the extremal black-hole case. Furthermore, in this zero-temperature case, we also demonstrate (even if only at the order-of-magnitude level) how a stringy correction arises in the gravity calculation.  For the Lifshitz black hole background, which is dual to the finite-temperature environment, the dissipative dynamics of the mirror become ohmic, and this gives a constant decoherence rate. This is consistent with the DSW effect in finite-temperature black holes. In this work, we also consider the decoherence of one of the EPR particles, in an EPR pair constructed holographically. The presence of the entangled partner suppresses the decoherence of the particle in the interference experiment when the two particles are in causal contact. When the particles in the EPR pair are out of causal contact, the decoherence pattern is not very different from that in the case with just one particle.

 The results in the current work imply that the holographic description of black holes is consistent with the perspective that black holes can be described by ordinary quantum systems. Our results are also consistent with the semiclassical description of black holes in studies of the DSW effect. This is perhaps unsurprising, given that we have used the large-$N$ holographic dictionary. In this respect, it will be interesting to study how finite-$N$ corrections change the decoherence patterns. Furthermore, the constant decoherence rate cannot continue indefinitely if the system in the interference experiment has more degrees of freedom than the black hole. The question is how decoherence terminates when we go beyond the semiclassical description of black holes. We do not know the answer yet, but a hint may come from our study of decoherence in holographic EPR pairs, where the decoherence of a particle due to the thermal environment is suppressed when it is causally connected to the other entangled particle.

\section*{Acknowledgments}

The work of D.S. L. was supported in part by the National Science and Technology Council, R.O.C. (NSTC 114-2112-M-259-007-MY3). The work of C.P. Y. was supported in part by the National Science and Technology Council, R.O.C. (NSTC 113-2112-M-259-009). C.P. Y. would also like to thank Chong-Sun Chu, Daine Danielson, and Norihiro Iizuka for useful comments on this work.

\end{document}